\documentstyle[preprint,aps]{revtex}

\def\G{\Gamma^{\vphantom{*}}}
\def\fr#1.#2.{{#1\over #2}}

\begin{document}
\tighten
\draft


\title{Fermion mixings vs $d=6$ proton decay}

\author{Pavel Fileviez P\'erez}

\address{~}

\address{{\it Max-Planck Institut f\" ur Physik \\
(Werner Heisenberg Institut) \\ F\" ohringer Ring 6.\\
80805 M\" unchen, Germany. \\ and}}
\address{{\it The Abdus Salam International Centre for
Theoretical Physics. \\ Strada Costiera 11, 34100.\\
Trieste, Italy}}

\maketitle

\begin{abstract}

It is well known, although sometimes ignored, that not only the $d=5$
but also $d=6$ proton decay depends on fermion mixings. In general we study
carefully the dependence of $d=6$ decay on fermion mixings 
using the effective operator approach. We find that without specifying
a theory it is impossible to make clear predictions. Even in a given
model, it is often not possible to determine all the physical
parameters. We point out that it is possible to make a clear test of 
any grand unified theory with symmetric Yukawa couplings.
We discuss in some detail realistic theories based on 
$SU(5)$ and $SO(10)$ gauge symmetry.
\end{abstract}

\section{Introduction}

The decay of the proton is the most dramatic prediction coming from
matter unification. Since the paper by Pati and Salam in 1973 \cite{PatiSalam},
proton decay has been the most important constraint for grand
unified theories 
\cite{SU(5),SUSYSU(5),SO(10),Weinberg,Zee,Sakai,Dimopoulos,Buras,Nathp1,Nathp2,Hisano1,Raby}. 
There are different operators contributing to the
nucleon decay in GUTs, in supersymmetric scenarios the $d=5$
contributions are the most important, but quite model dependent. They 
depend on the whole SUSY spectrum, on the structure of the Higgs
sector and on fermion masses. In recent years these contributions
have been under discussion, in order to understand if the minimal
supersymmetric $SU(5)$ \cite{SU(5),SUSYSU(5)} is ruled out
\cite{Murayama,Goto}. There are several solutions to this 
very important issue in the context of the
minimal SUSY SU(5) \cite{Bajc1,Bajc2}. 
\\
\\
The $d=6$ contributions for proton decay in general are the
second more important, but they are less model dependent. 
From the non-diagonal part of the gauge field we get 
the gauge contributions, which basically depend 
only on fermion masses. The remaining $d=6$ operators coming from 
the Higgs sector are less important and they are quite model dependent, 
since we can have different structures in the Higgs sector. There are
several models where due to a specific structure of the Higgs sector,
the $d=5$ operators contributing to the decay of the proton are not 
present\cite{Babu}.
\\
\\
In general we study in detail the gauge $d=6$ contributions. 
Assuming that in the future the decay of the proton will be 
measured, we analyze all 
possible information that we could get from these experiments. Using
this information we will study the possibility to test the realistic 
grand unified theories based on the $SU(5)$ and $SO(10)$ gauge groups. 
Our analysis is valid in supersymmetric and non-supersymmetric GUT scenarios.

\section{d=6 operators}
%
Using the properties of the Standard Model fields we can write down 
the possible $d=6$ operators contributing to the decay of the proton, which are
$SU(3)_C \times SU(2)_L \times U(1)_Y$ invariant \cite{Weinberg,Zee,Sakai}:
\begin{eqnarray}
\label{O1}
\textit{O}^{B-L}_I&=& k^2_1
\ \epsilon_{ijk} \ \epsilon_{\alpha \beta} 
\ \overline{u_{i a}^C} \ \gamma^{\mu} \ Q_{j \alpha a}   \
\overline{e_b^C} \ \gamma_{\mu} \ Q_{k \beta b}\\ 
\label{O2}
\textit{O}^{B-L}_{II}&=& k^2_1
\ \epsilon_{ijk} \ \epsilon_{\alpha \beta}
\ \overline{u_{i a}^C} \ \gamma^{\mu} \ Q_{j \alpha a}   \
\overline{d^C_{k b}} \ \gamma_{\mu} \ L_{\beta b} \\
\label{O3}
\textit{O}^{B-L}_{III}&=& k^2_2
\ \epsilon_{ijk} \ \epsilon_{\alpha \beta}
\ \overline{d_{i a}^C} \ \gamma^{\mu} \ Q_{j \beta a}   \
\overline{u_{k b}^C} \ \gamma_{\mu} \ L_{\alpha b} \\
\label{O4}
\textit{O}^{B-L}_{IV}&=& k^2_2
\ \epsilon_{ijk} \ \epsilon_{\alpha \beta}
\ \overline{d^C_{i a}} \ \gamma^{\mu} \ Q_{j \beta a}   \
\overline{\nu_b^C} \ \gamma_{\mu} \ Q_{k \alpha b}
\end{eqnarray}
In the above expressions $k_1= g_{GUT} M^{-1}_{(X,Y)}$, and 
$k_2= g_{GUT} {M^{-1}_{(X^{'},Y^{'})}}$, 
where $M_{(X,Y)}$, $M_{(X^{'},Y^{'})}$ $\sim M_{GUT}\approx
10^{16} \ GeV$ and $g_{GUT}$ are the masses of the superheavy gauge bosons
and the coupling at the GUT scale. $Q= ( u, d)$, $L= ( \nu, e)$; 
$i$, $j$ and $k$ are the color indices, $a$ and $b$ are the family
indices, and $\alpha, \beta =1,2$. 
\\
The effective operators $\textit{O}^{B-L}_I$ and
$\textit{O}^{B-L}_{II}$ (eqs. \ref{O1} and \ref{O2}) 
appear when we integrate out the superheavy gauge fields $(X, Y)=(3,2,5/3)$,
where the $X$ and $Y$ fields have electric charge $4/3$ and $1/3$ 
respectively. This is the case in theories based on the gauge group
SU(5). Integrating out $(X^{'}, Y^{'})=(3,2,-1/3)$ we obtain the 
operators $\textit{O}^{B-L}_{III}$ and $\textit{O}^{B-L}_{IV}$ 
(eqs. \ref{O3} and \ref{O4}), the electric charge of $Y^{'}$ is $-2/3$, while 
$X^{'}$ has the same charge as $Y$. In $SO(10)$ theories all these
superheavy fields are present. 
Notice that all these operators conserve $B-L$, i.e. the proton always
decays into an antilepton. A second selection rule $\Delta S/
\Delta B= -1,0$ is satisfied for those operators \cite{Langacker}.
\\
\\
Using the operators listed above, we can write the effective operators
for each decay channel in the physical basis:

\begin{eqnarray}
\label{Oec}
\textit{O}(e_{\alpha}^C, d_{\beta})&=& k_1^2 \ c(e^C_{\alpha},
d_{\beta}) \ \epsilon_{ijk} 
\ \overline{u^C_i} \ \gamma^{\mu} \ u_j \ \overline{e^C_{\alpha}} \
\gamma_{\mu} \ d_{k \beta} \\
\label{Oe}
\textit{O}(e_{\alpha}, d^C_{\beta})&=& c(e_{\alpha}, d^C_{\beta}) \ \epsilon_{ijk} \ 
\overline{u^C_i} \ \gamma^{\mu} \ u_j \ \overline{d^C_{k \beta}} \
\gamma_{\mu} \ e_{\alpha}\\
\label{On}
\textit{O}(\nu_l, d_{\alpha}, d^C_{\beta} )&=& c(\nu_l, d_{\alpha}, d^C_{\beta}) \
\epsilon_{ijk} \ \overline{u^C_i} \ \gamma^{\mu} \ d_{j \alpha}
\ \overline{d^C_{k \beta}} \ \gamma_{\mu} \ \nu_l \\
\label{OnC}
\textit{O}(\nu_l^C, d_{\alpha}, d^C_{\beta} )&=& k_2^2 \ 
c(\nu_l^C, d_{\alpha}, d^C_{\beta}) \
\epsilon_{ijk} \
\overline{d_{i \beta}^C} \ \gamma^{\mu} \ u_j \ \overline{\nu_l^C} \
\gamma_{\mu} \ d_{k \alpha}
\end{eqnarray}
\\
\\
\\
\\
where:
\begin{eqnarray}
\label{cec}
c(e^C_{\alpha}, d_{\beta})&=& V^{11}_1 V^{\alpha \beta}_2 + ( V_1 V_{UD})^{1
\beta}( V_2 V^{\dagger}_{UD})^{\alpha 1} \\
\label{ce}
c(e_{\alpha}, d_{\beta}^C) &=& k^2_1  \ V^{11}_1 V^{\beta \alpha}_3 +  \ k_2^2 \
(V_4 V^{\dagger}_{UD} )^{\beta 1} ( V_1 V_{UD} V_4^{\dagger} V_3)^{1 \alpha}\\
\label{cnu} 
c(\nu_l, d_{\alpha}, d^C_{\beta})&=& k_1^2 \ ( V_1 V_{UD} )^{1 \alpha}
( V_3 V_{EN})^{\beta l} + \ k_2^2 \ V_4^{\beta \alpha}( V_1 V_{UD}
V^{\dagger}_4 V_3 V_{EN})^{1l} \\
\label{cnuc}
c(\nu_l^C, d_{\alpha}, d^C_{\beta})&=& ( V_4 V^{\dagger}_{UD} )^{\beta
 1} ( U^{\dagger}_{EN} V_2)^{l \alpha }+ V^{\beta \alpha}_4
 (U^{\dagger}_{EN} V_2 V^{\dagger}_{UD})^{l1};  \ \ \ \alpha = \beta
 \neq 2
\end{eqnarray}
The mixing matrices $V_1= U_C^{\dagger} U$, $V_2=E_C^{\dagger}D$,
$V_3=D_C^{\dagger}E$, $V_4=D_C^{\dagger} D$, $V_{UD}=U^{\dagger}D$,
$V_{EN}=E^{\dagger}N$ and $U_{EN}= {E^C}^{\dagger} N^C$.   
The quark mixings are given by $V_{UD}=U^{\dagger}D=K_1 V_{CKM} K_2$,
where $K_1$ and $K_2$ are diagonal matrices containing three and two
phases respectively. The leptonic mixing $V_{EN}=K_3 V^D_l K_4$ in case
of Dirac neutrino, or $V_{EN}=K_3 V^M_l$ in the Majorana case, $V^D_l$
and $V^M_l$ are the leptonic mixings at low energy in the Dirac and
Majorana case respectively.
\\
\\
Notice that in general to predict the lifetime of the proton
due to the presence of $d=6$ operators we have to know 
$k_1$, $k_2$, $V^{1b}_1$, $V_2$, $V_3$, $V_4$ and $U_{EN}$. In addition
we have three diagonal matrices containing CP violation phases,
$K_1$, $K_2$ and $K_3$, in the case that the neutrino is Majorana.
In the Dirac case there is an extra matrix with two more phases.

\section{Two bodies decay channels of the nucleon}
As we know the gauge $d=6$ operators conserve $B-L$, 
therefore the nucleon decays into a meson and an antilepton.  
Let us analyze all different channels. Assuming that
in the proton decay experiments \cite{exp} one can not 
distinguish the flavour of the neutrino and the 
chirality of charged leptons in the exit
channel, and using the chiral Lagrangian techniques 
(see reference \cite{Chadha}), the decay rate of the different channels due to 
the presence of the gauge $d=6$ operators are given by:

\begin{eqnarray}
\label{A1}
\Gamma(p \to K^+\bar{\nu})
		&=& \frac{(m_p^2-m_K^2)^2}{8\pi m_p^3 f_{\pi}^2} A_L^2 
\left|\alpha\right|^2
\sum_{i=1}^3 \left|\frac{2m_p}{3m_B}D \ c(\nu_i, d, s^C) 
+ [1+\frac{m_p}{3m_B}(D+3F)] c(\nu_i,s, d^C)\right|^2\\
\label{A2}
\Gamma(p \to \pi^+\bar{\nu})
		&=&\frac{m_p}{8\pi f_{\pi}^2}  A_L^2 \left|\alpha
		\right|^2 (1+D+F)^2 
\sum_{i=1}^3 \left| c(\nu_i, d, d^C) \right|^2\\
\label{A3}
\Gamma(p \to \eta e_{\beta}^+) 
		&=& {(m_p^2-m_\eta^2)^2\over 48 \pi f_\pi^2 m_p^3}
A_L^2 \left|\alpha \right|^2 (1+D-3 F)^2 \{ \left| c(e_{\beta},d^C)\right|^2 + k_1^4 \left|c(e^C_{\beta}, d)\right|^2 \}\\
\label{A4}
\Gamma (p \to K^0 e_{\beta}^+) 
		&=& {(m_p^2-m_K^2)^2\over 8 \pi f_\pi^2 m_p^3}  A_L^2
		\left|\alpha\right|^2 [1+{m_p\over m_B} (D-F)]^2 
\{ \left|c(e_{\beta},s^C)\right|^2 + k_1^4 \left|c(e^C_{\beta},s)\right|^2\}\\
\label{A5}
\Gamma(p \rightarrow \pi^0 e_{\beta}^+)
           &=& \frac{m_p}{16\pi f_{\pi}^2} A_L^2 \left|\alpha\right|^2
		(1+D+F)^2 \{ \left|c(e_{\beta},d^C)\right|^2 + k_1^4
		\left|c(e^C_{\beta},d)\right|^2 \}
\end{eqnarray}
\newpage 
\begin{eqnarray}
\label{A6}
\G(n \to K^0 \overline\nu)&=& 
\fr (m_n^2-m_K^2)^2. 8 \pi m_n^3 f_\pi^2. A_L^2 \left|\alpha\right|^2 \times
\nonumber\\ &&\times  \sum_{i=1}^3 \left|c(\nu_i,d,s^C) [1+\fr m_n.3 m_B. (D-3 F)]-c(\nu_i,s,d^C)[1+\fr
		m_n.3 m_B.(D+3 F)]\right|^2\\
\label{A7}
\G(n \to \pi^0 \overline\nu)&=&\fr m_n .16 \pi f_\pi^2. A_L^2
		\left|\alpha\right|^2 (1+D+F)^2 \sum_{i=1}^3 \left|c(\nu_i, d,
		d^C)\right|^2\\
\label{A8}
\G(n \to \eta \overline\nu)&=& \fr (m_n^2-m_\eta^2)^2. 48 \pi m_n^3
f_\pi^2. A_L^2 \left|\alpha\right|^2 (1+D-3 F)^2 \sum_{i=1}^3 \left|c(\nu_i, d, d^C)\right|^2\\
\label{A9}
\G(n \to \pi^- e^+_{\beta})&=&\fr m_n. 8 \pi f_\pi^2. A_L^2
		\left|\alpha\right|^2 (1+D+F)^2
		\{\left|c(e_{\beta}, d^C)\right|^2 + k_1^4 \left|c(e^C_{\beta},d)\right|^2\}
\end{eqnarray} 
In the above equations $m_B$ is an average Baryon mass satisfying $m_B \approx
m_\Sigma \approx m_\Lambda$, $D$, $F$ and $\alpha$ are the parameters
of the chiral lagrangian, and all other notation follows
\cite{Chadha}. Here all coefficients of four-fermion operators are evaluated at
$M_Z$ scale. $A_L$ takes into account renormalization from $M_Z$ to 1
GeV. $\nu_i= \nu_e, \nu_{\mu}, \nu_{\tau}$ and $e_{\beta}= e, \mu$. 
\\
\\
Let us analyze all different channels. When the nucleon decays into 
a strange meson plus an antineutrino the
amplitudes (eqs. \ref{A1} and \ref{A6}) of these channels are
proportional to a linear combination of the 
coefficients $c(\nu_i, s, d^C)$ and $c(\nu_i, d, s^C)$ . 
In the case of the nucleon decays into a light unflavored
meson plus an antineutrino, the amplitudes (eqs. \ref{A2}, \ref{A7} and \ref{A8}) 
are proportional to $\sum_{i=1}^{3} c(\nu_i, d, d^C)$. Looking at the channels 
with a charged antilepton, we see that the amplitudes (eqs. \ref{A3}, \ref{A5}
and \ref{A9}) of the channels with a light meson are proportional to a
linear combination of the coefficients $c(e_\alpha, d^C)$ and $c(e^C_\alpha,
d)$, while in the case that we have a strange meson they 
are proportional to a linear combination of $c(e_\alpha,
s^C)$ and $c(e^C_\alpha, s)$ (eq. \ref{A4}). 
If the neutrinos are Dirac-like we have extra channels 
to the decay of the nucleon, where we
have the decays into $\nu^C_i$ and a meson. The amplitudes in this case
are proportional to $c(\nu_i^C, d, d^C)$ , $c(\nu^C_i, s, d^C)$ and
$c(\nu_i^C, d, s^C)$ respectively. Notice that from the 
radiative decays \cite{radiative} we get the same information 
as in the case of the decays into a charged antilepton.
\\ 
Note that from the equations \ref{A1}-\ref{A9} we can get only seven
relations for all coefficients of the gauge d=6 operators contributing
to nucleon decay. Therefore, if we want to test a grand unified theory the
number of physical quantities entering in the proton decay amplitude
must be less than seven. This is an important result which helps us to
know when it is possible to test a GUT scenario. However, as we will
see in the next section looking only at the antineutrino channels we can
get interesting predictions. 
%
\section{testing GUT models}
%
Let us analyze the possibility to test the realistic grand unified
models, the $SU(5)$ and $SO(10)$ theories respectively. Let us make an
analysis of the operators in each theory, and study the physical
parameters entering in the predictions for proton decay.
Here we do not assume any particular model for fermion masses,
in order to be sure that we can test the grand unification idea.
\\
\\
In these models the diagonalization of the Yukawa matrices is given by:
\begin{eqnarray}
\label{YUd}
U^T_C \ Y_U \ U &=& Y_U^{diag}\\
\label{YDd}
D^T_C \ Y_D \ D &=& Y_D^{diag}\\
\label{YEd}
E^T_C \ Y_E \ E &=& Y_E^{diag}
\end{eqnarray}

\begin{itemize}
\item a GUT based on $SU(5)$ 
\end{itemize}
Let us start with the simplest grand unified theories, 
which are based on the gauge group $SU(5)$. 
In these theories the unification of quark and leptons is realized in two
irreducible representations, $10$ and $\overline{5}$. The minimal Higgs
sector is composed by the adjoint representation $\Sigma$, and two Higgses
$5_H$ and $\overline{5}_H$ in the fundamental and anti-fundamental
representations respectively \cite{SU(5),SUSYSU(5)}, if we want to
keep the minimal Higgs sector and write down a realistic $SU(5)$
theory, we need to introduce non-renormalizable operators, Planck 
suppressed operators, to get the correct quark-lepton mass relations. A
second possibility is introduce a Higgs in the $45_H$ representation. 
\\
In this case we have only the operators $\textit{O}^{B-L}_I$ (eq. 1), 
and $\textit{O}^{B-L}_{II}$ (eq. 2) contributing to the decay of the
proton. Using the equations \ref{cec}-\ref{cnuc}, and taking 
$k_2 \equiv 0$ the coefficients for the proton decay predictions are
given by:

\begin{eqnarray}
\label{cecSU5}
c(e^C_{\alpha}, d_{\beta})_{SU(5)}&=& V^{11}_1 V^{\alpha \beta}_2 + ( V_1 V_{UD})^{1
\beta}( V_2 V^{\dagger}_{UD})^{\alpha1} \\
\label{ceSU5}
c(e_{\alpha}, d_{\beta}^C)_{SU(5)} &=& k^2_1  \ V^{11}_1 V^{\beta \alpha}_3 \\ 
\label{cnuSU5} 
c(\nu_{l}, d_{\alpha}, d^C_{\beta})_{SU(5)}&=&  k_1^2 \ ( V_1 V_{UD} )^{1 \alpha}
( V_3 V_{EN} )^{\beta l};  \ \ \ \alpha = \beta
 \neq 2\\
\label{cnucSU5}
O(\nu_b^C, d_{\alpha}, d^C_{\beta})_{SU(5)}&=& 0.
\end{eqnarray}
We see from these expressions that in order to make predictions 
in any theory based on the $SU(5)$ gauge group using proton decay, 
we have to know $k_1$, $V^{1i}_1$ and the matrices $V_2$, and $V_3$. 
Note that in a SU(5) theory there are not decays into a $\nu^C$, even
if the neutrino is a Dirac-like particle (see eq. \ref{cnucSU5}), it
could be a possibility to distinguish a $SU(5)$ theory of the rest of GUTs.
\\
In order to compute the decay rate into antineutrinos we must use the
following relation:
\begin{eqnarray}
\label{sumSU(5)}
\sum_{l=1}^3 c(\nu_l, d_{\alpha}, d_{\beta})_{SU(5)}^* c(\nu_l, d_{\gamma},
d_{\delta})_{SU(5)}&=& k_1^4 (V_1^* K_1^* V_{CKM}^*)^{1 \alpha}
(K_2^*)^{\alpha \alpha} (V_1 K_1 V_{CKM})^{1 \gamma} K_2^{\gamma
\gamma} \delta^{\beta \delta}
\end{eqnarray}
Using this expression we can see that the antineutrino channel depends on the matrices $V_1$
and $K_1$. Since we have only three independent equations (eqs. \ref{A1},
\ref{A2} and \ref{A6}) for these channels, it is clear that we can not test a GUT 
theory based on $SU(5)$. Notice that from the channels with charged
leptons is even more difficult to get some information, due to the
presence of the matrices $V_2$, $V_3$ and the elements
$V_1^{1i}$. In the naive case without all CP violation sources beyond
$V_{CKM}$ we could get the information about $V_1$ from the nucleon
decays into antineutrinos.
\\
\\ 
Let us analyze a particular case, the unrealistic minimal $SU(5)$ model, where
$Y_U=Y_U^T$ and $Y_D=Y_E^T$ (see reference \cite{Mohapatra1}), 
in this case we have the following relations:

\begin{eqnarray}
c(e^C_{\alpha}, d_{\beta})^{unreal-min}_{SU(5)}&=& (K_u^*)^{11} [
\delta^{\alpha \beta} +  V_{CKM}^{1 \beta} K_2^{\beta \beta}
(K_2^{*})^{\alpha \alpha} (V_{CKM}^{\dagger})^{\alpha 1}]\\
\label{ceSU5}
c(e_{\alpha}, d_{\beta}^C)^{unreal-min}_{SU(5)} &=& k^2_1  \ (K_u^*)^{11} \delta^{\beta \alpha}\\ 
\label{cnuSU5} 
c(\nu_{l}, d_{\alpha}, d^C_{\beta})^{unreal-min}_{SU(5)}&=&  k_1^2 \
(K_u^*)^{11} K_1^{11} V_{CKM}^{1 \alpha} K_2^{\alpha \alpha}
V_{EN}^{\beta l};  \ \ \ \alpha = \beta
 \neq 2.
\end{eqnarray}
\\
Notice that in this naive GUT model, all the channels are determined by
$V_{CKM}$. Unfortunately it is a prediction that we lost in the case
of realistic versions of $SU(5)$. However, if this modification of the
theory does not change the relation $Y_U=Y_U^T$, we could test a
$SU(5)$ theory from the nucleon decays into an antineutrino (see equation \ref{sumSU(5)}).
\\
\begin{itemize}
\item a GUT model with symmetric Yukawa couplings.
\end{itemize}
There are many examples of grand unified theories with symmetric 
Yukawa couplings. This is the case of SO(10)\cite{SO(10)} 
theories with two Higgses $10_H$ and $126_H$, including the minimal
supersymmetric SO(10) model\cite{RSO(10)1,RSO(10)2}.  
\\
\\
In reference \cite{DeRujula} has been investigated the dependence
of the $d=6$ gauge contributions on fermion mixings. They consider two
different cases, the naive minimal $SO(10)$, where all
fermion masses arise from Yukawa couplings to $10_H$, and 
the case where we have the Higgses
$10_H$ and $126_H$. Assuming only two generations, and neglecting the
possible mixings which appear when the neutrino mass matrix 
is diagonalized, they showed approximately 
that the predictions for the decay channels
$p \to \pi^+ \bar{\nu}$ and $p \to K^0 l^+$ do not change 
in the different models for fermion masses. At the same time, it has
been showed that the predictions for the decays $p \to K^0 e^+$, and
$p \to \mu^+ \pi$ are quite different in these two scenarios for
fermion masses. 
\\
\\
In this section we will analyze the
properties of all decays in those theories, using the fact that the Yukawa
matrices are symmetric. We will take into account the mixings of the
third generation and all possible CP violation effects.
\\
\\
In theories with symmetric Yukawa couplings we get 
the following relations for the mixing matrices, $U_C = U K_u$, $D_C = D K_d$
and $E_C = E K_e$, where $K_u$, $K_d$ and $K_e$ are diagonal matrices
containing three CP violating phases. In those cases 
$V_1 = K^*_u$, $V_2= K^*_e V^{\dagger}_{DE}$, $V_3=K^*_d V_{DE}$ and 
$V_4=K^*_d$. Using these relations the coefficients in equations 
\ref{cec}-\ref{cnuc} are given by:

\begin{eqnarray}
\label{cecs}
c(e^C_{\alpha}, d_{\beta})_{sym}&=& (K_u^*)^{11}(K_e^*)^{\alpha
\alpha}
[ \delta^{\beta i} + V^{1 \beta}_{CKM} K_2^{\beta \beta} (K_2^*)^{ii}
(V^{\dagger}_{CKM})^{i1}  ] (V^*_{DE})^{i \alpha}\\
\label{ces}
c(e_{\alpha}, d_{\beta}^C)_{sym} &=& (K^*_u)^{11}(K^*_d)^{\beta \beta} [
k_1^2 \delta^{\beta i} + k_2^2 (K_2^*)^{\beta
\beta}(V^{\dagger}_{CKM})^{\beta 1} V^{1i}_{CKM} K_2^{ii} ](V^{i \alpha}_{DE})\\
\label{cnus} 
c(\nu_l, d_{\alpha}, d^C_{\beta})_{sym}&=& (K_u^*)^{11}K_1^{11}
[ k_1^2  \delta^{\alpha i} \delta^{\beta j} +
k_2^2 \delta^{\alpha \beta} \delta^{ij} (K_d^*)^{\alpha \alpha}
K_d^{ii}](V_{CKM} K_2)^{1i} (K_d^* V_{DE} V_{EN})^{jl}\\
\label{cnucs}
c(\nu_l^C, d_{\alpha}, d^C_{\beta})_{sym}&=& (K^*_d)^{\beta
\beta}(K_1^*)^{11} [ (K^*_2)^{\beta \beta} (V^{\dagger}_{CKM})^{\beta
1} \delta^{\alpha i} + \delta^{\alpha \beta} (K_2^*)^{ii}
(V^{\dagger}_{CKM})^{i1}](U_{EN}^{\dagger} K_e^*
V_{DE}^{\dagger})^{li}
\end{eqnarray}
with $\alpha = \beta \neq 2\nonumber$.
\\
\\
\\
Notice all overall phases in the different coefficients. In order to
compute the decay rate into an antineutrino we need the following expression:

\begin{eqnarray}
\sum_{l=1}^3 c(\nu_l, d_{\alpha}, d_{\beta})_{sym}^* c(\nu_l, d_{\gamma},
d_{\delta})_{sym}&=&[ k_1^2 \delta^{\alpha i} \delta^{\beta j} + k_2^2
\delta^{\alpha \beta} \delta^{ij} K_d^{\alpha \alpha}
(K_d^*)^{ii}]\nonumber\\
&&[ k_1^2 \delta^{\gamma i^{'}} \delta^{\delta j} + k_2^2 \delta^{\gamma
\delta} \delta^{i^{'} j}(K_d^*)^{\gamma \gamma} K_d^{i^{'} i^{'}}
](V^*_{CKM} K_2^*)^{1i} (V_{CKM} K_2)^{1i^{'}}\nonumber\\
\end{eqnarray}
Using the above expression, and equation \ref{A1} we find that it
is possible to determine the factor $k_1= g_{GUT}/M_{(X,Y)}$:

\begin{eqnarray}
\label{k1}
k_1&=& \frac{Q_1^{1/4}}{[ \left|A_1\right|^2
\left|V_{CKM}^{11}\right|^2+ \left|A_2\right|^2
\left|V_{CKM}^{12}\right|^2]^{1/4} } 
\end{eqnarray} 
where:
\begin{eqnarray}
Q_1&=& \frac{8 \pi m_p^3 f^2_{\pi} \Gamma(p \to K^+ \bar{\nu})}
{(m_p^2 - m_K^2)^2 A_L^2 \left|\alpha \right|^2}\\
A_1&=& \frac{2 m_p }{3 m_B} D\\
A_2&=& 1 + \frac{m_p}{3 m_B}(D + 3F)
\end{eqnarray}
Notice that we have an expression for $k_1$, which is independent of
the unknown mixing matrices and the CP violating phases. 
In other words, we find that the 
amplitude of the decay $p \to K^+ \bar{\nu}$ is independent of all
unknown mixings and CP violating phases, this 
only depends on the factor $k_1$. Therefore it is a possibility to test
any grand unified theory with symmetric Yukawa matrices through this channel. 
\\
\\
Once we know $k_1$, and using the expression \ref{A2} we can find the 
factor $k_2$, solving the following equation:

\begin{eqnarray}
\label{k2}
k_2^4 + 2 k_2^2 k_1^2 \left|V_{CKM}^{11}\right|^2 + k_1^4
\left|V^{11}_{CKM}\right|^2 - \frac{8 \pi f^2_{\pi} \Gamma(p \to \pi^+ \bar{\nu})}
{m_p A_L^2 \left|\alpha \right|^2 (1+D+F)^2} &=&0
\end{eqnarray}
\begin{eqnarray}
\label{k2s}
k_2 &=&k_1 \left|V_{CKM}^{11}\right| \{ - 1 + \sqrt{Q_2} \}^{1/2} 
\end{eqnarray}
with:
\begin{eqnarray}
Q_2&=& 1 + \frac{8 \pi f_{\pi}^2 \Gamma( p \to \pi^+ \bar{\nu})}{
k_1^4 \left|V_{CKM}^{11}\right|^4 m_p A^2_L \left|\alpha\right|^2 (1+D+F)^2}- \left|V_{CKM}^{11}\right|^{-2}
\end{eqnarray}
Using the condition $ Q_2 > 1$, we get the following relation:
\begin{eqnarray}
\label{R1}
\frac{\tau(p \to K^+
\bar{\nu})}{\tau(p \to \pi^+ \bar{\nu})}&>&\frac{m_p^4 \left|V_{CKM}^{11}\right|^2 (1+D+F)^2}{(m_p^2 -m_K^2)^2 [\left|A_1\right|^2 \left|V_{CKM}^{11}\right|^2+ \left|A_2\right|^2 \left|V_{CKM}^{12}\right|^2 ]}
\end{eqnarray}
It is a clear prediction of a GUT model with symmetric Yukawa couplings. 
\\
\\
Using the expressions \ref{A1}, \ref{A2}, \ref{A6}, \ref{A7}, and 
\ref{A8} we can get the following relations:
\begin{eqnarray}
\label{R2}
\frac{\tau(n \to K^0 \bar{\nu})}{\tau(p \to K^+ \bar{\nu})}&=&
\frac{m_n^3 (m_p^2 -m_K^2)^2 [\left|A_1\right|^2
\left|V_{CKM}^{11}\right|^2+ \left|A_2\right|^2
\left|V_{CKM}^{12}\right|^2 ]}{m_p^3 (m_n^2 -m_K^2)^2 [\left|A_3
\right|^2 \left|V_{CKM}^{11}\right|^2+ \left|A_2\right|^2
\left|V_{CKM}^{12}\right|^2 ]}\\
\label{R3}
\frac{\tau(n \to \pi^0 \bar{\nu})}{\tau(p \to \pi^+ \bar{\nu})}&=&
\frac{2 m_p}{m_n}\\
\label{R4}
\frac{\tau(n \to \eta^0 \bar{\nu})}{\tau(p \to \pi^+ \bar{\nu})}&=&
\frac{6 m_p m_n^3 (1+D+F)^2}{(m_n^2-m_{\eta}^2)^2 (1-D-3F)^2}
\end{eqnarray}
with:
\begin{eqnarray}
A_3= 1 + \frac{m_n}{3 m_B}(D-3F)
\end{eqnarray}
Notice that using the expressions for $k_1$ and $k_2$ (eqs. \ref{k1} and
\ref{k2s}), and the relation between the different decay rates of the
neutron and the proton into an antineutrino (eqs. \ref{R1}-\ref{R4}), 
we can conclude that it is possible to make a clear test of a 
grand unified theory with symmetric Yukawa couplings.  
\\
\\
As we say before, there are realistic $SO(10)$ theories 
with symmetric Yukawa couplings. In a $SO(10)$ theory all fermions 
of a family live in the $16_F$ spinor representation\cite{SO(10)}. 
In this case the coefficients for the gauge d=6 operators are given by
the equations \ref{cec}-\ref{cnuc}.
\\
\\
Let us analyze the most realistic and studied SO(10) theories, where
all Yukawa couplings are symmetric. It is the case of theories with the
$10_H$ and/or $126_H$ Higgses\cite{RSO(10)1,RSO(10)2,Charan1,RSO(10)3,Charan,FM1,FM2,FM3,Fukuyama1,Fukuyama2}. 
We have already studied the case of GUT
models with symmetric Yukawa couplings, where we pointed out the
possibility to make a consistent check of these theories.  
In order to predict the decay rates into charged antileptons in this
case, we have to know the matrices $K_2$ and $V_{DE}$ (see eqs. \ref{cecs}
and \ref{ces}). In those $SO(10)$ theories there is a specific expression 
for the matrix $V_{DE}$: 
\begin{eqnarray}
\label{VDE}
4 V^T_{UD} K^*_u Y^{diag}_U V_{UD}-(3 \tan \alpha_{10} + \tan
\alpha_{126}) K^*_d Y^{diag}_D&=& V^*_{DE}K^*_e Y^{diag}_E
V^{\dagger}_{DE} ( \tan \alpha_{10}- \tan \alpha_{126})
\end{eqnarray}
In the above expressions $\tan \alpha_{10}= v^{U}_{10} / v^{D}_{10}$,
and $\tan \alpha_{126}= v^{U}_{126} / v^{D}_{126}$. In equation
\ref{VDE} we see explicitly the relation between the different factors
entering in the proton decay predictions. 
\\
\\
To compute the amplitude for proton decay into charged antileptons we
need the following expression: 

\begin{eqnarray}
\label{CL}
\sum_{\alpha =1}^2 c(e^C_{\alpha}, d_{\beta})^*_{sym}
c(e^C_{\alpha}, d_{\gamma})_{sym}
&=& 
[ \delta^{\beta i} + V^{1 \beta}_{CKM} K_2^{\beta \beta} (K_2^*)^{ii}
(V^{\dagger}_{CKM})^{i1}  ] \nonumber\\
&&[ \delta^{\gamma j} + V^{1 \gamma}_{CKM} K_2^{\gamma \gamma} (K_2^*)^{jj}
(V^{\dagger}_{CKM})^{j1} ] 
\sum^2_{i=1} V^{i \alpha}_{DE} (V^{j \alpha}_{DE})^{*}
\end{eqnarray}
Therefore the amplitude of the channels with charged antileptons 
always depend on the matrices $K_2$ and $V_{DE}$. 
Therefore it is not possible to make a
clear test of the theory through those channels, they are useful to
distinguish between different models for fermion masses with symmetric Yukawa
matrices. Notice that in reference \cite{DeRujula} has been showed 
that the predictions for the decay channel $p \to l^+ K^0$ are the
same in different models for fermion masses, however as we can
appreciate from equation \ref{CL} it is not true in the general case
when we consider all generations and the extra CP violating phases.

\section{Conclusions}
We have studied in detail the predictions coming from 
the gauge $d=6$ operators, the less model dependent 
contributions for proton decay.
Analyzing the different decay channels, we find 
that there are only seven independent equations for the coefficients 
involved in the two bodies decay channels for proton decay. In general 
we could say that the number of physical parameters involved in those
predictions must be less than seven. 
\\
We have pointed out that it is possible to make a clear test of 
any grand unified theory with symmetric Yukawa couplings through the
decay of the nucleon, 
since in these cases the decay rates of 
the nucleon into an antineutrino are independent of 
the mixings matrices and the new sources of CP
violation beyond $V_{CKM}$ and $V_l$, they depend only on the factors
$k_1$ and $k_2$. The relations between the decays of the proton and
the neutron into an antineutrino have been found. Notice that it is the case 
of realistic grand unified theories based on the $SO(10)$ 
gauge group. The predictions for the decay channels with charged
leptons are not the same in different models for fermion masses
with symmetric Yukawa couplings, therefore they could be useful to
distinguish between different models.  
Our results are valid in supersymmetric and
non-supersymmetric scenarios. 
\newpage
\acknowledgments

We would like to thank Goran Senjanovi\'c and Borut Bajc for
invaluable discussions and comments. We would like to thank 
Ariel Garcia, Lotfi Boubekeur and Brett Viren for comments. 
We would like to thank the High Energy Section of the ICTP 
for hospitality and support.

\end{document}